\documentclass{aa}
\usepackage{graphicx}
\usepackage{txfonts}
\usepackage{hyperref}
\begin{document}

   \title{High reddening patches in \textit{Gaia} DR2}

   \subtitle{Possible artifacts or indication of star formation at the edge of the Galactic disk}
   \author{Leire Beitia-Antero\thanks{L. Beitia-Antero: lbeitia@ucm.es}
          \inst{1,2}
          \and
          Ana In\'es G\'omez de Castro \inst{1}
          \and
          Ra\'ul de la Fuente Marcos \inst{1}
          }

   \institute{AEGORA Research Group, Facultad de CC. Matemáticas, Universidad
     Complutense de Madrid
         \and
         Departamento de F\'isica de la Tierra y Astrof\'isica,
         Facultad de CC. F\'isicas, Universidad Complutense de Madrid.
             }

   \date{Received XX YY, ZZZZ; accepted XX YY, ZZZZ}

   \titlerunning{High reddening patches in \textit{Gaia} DR2}
 
  \abstract
      {Deep GALEX UV data show that the extreme outskirts of some spiral galaxies are teeming with star formation.
Such young stellar populations evolving so far away from the bulk of their host galaxies challenge our overall understanding of how star
formation proceeds at galactic scales. It is at present unclear whether our own Milky Way may also exhibit ongoing and recent star
formation beyond the conventional edge of the disk ($\sim 15$ kpc).}
      {Using \textit{Gaia} DR2 data, we aim to determine if such a
        population is present in the Galactic halo, beyond the nominal
        radius of the Milky Way disk.}
      {We  studied the kinematics of \textit{Gaia} DR2 sources
        with parallax values between 1/60 and 1/30 milliarcseconds towards
        two regions that show abnormally high values of extinction and
        reddening; the results are compared with predictions
        from GALAXIA Galactic model. We   also plotted the color--magnitude
        (CM) diagrams with heliocentric
        distances computed inverting the parallaxes, and  
        studied the effects of the large parallax errors by Monte
      Carlo sampling.}
      {The kinematics point towards a Galactic origin for one of the regions, while the provenance of the
        stars in the other   is not clear. A spectroscopic analysis of some of the sources in the
        first region confirms that
        they are located in the halo. The CM diagram
        of the sources suggests that some of them are young.}
   {}


   \maketitle
   
%
\section{Introduction}

It has been customarily assumed that star formation cannot proceed
within low gas density environments, such as those that characterize 
the outermost regions of galactic disks \citep{Ferguson2002,Bianchi2009}.
However, the Galaxy Ultraviolet Explorer (GALEX, \citealt{Martin2005})
revealed the presence of young stellar populations 
in the extreme outskirts of spiral galaxies, well beyond where the
bulk of the galactic-scale star formation was assumed to take 
place \citep{Bianchi2005,Thilker2005,Thilker2007}. A common feature of all these galaxies is that they lie in
groups, as do most of the galaxies in the Local Universe, including the Milky Way,  and hence  extended ultraviolet
emission could be a natural consequence  \citep{Marino2010}. Nevertheless, 
the presence of young massive stars in the outskirts of our
Galaxy is very difficult to confirm in the ultraviolet range because of the 
high extinction areas close to the Galactic plane \citep{Schultheis2015,Schlafly2016,Schlafly2017}.
It is therefore necessary
to follow an alternative approach and look for young stars in other
spectral ranges. \par
With the second data release of the \textit{Gaia}
mission (\textit{Gaia} DR2, \citealt{GaiaMission,GaiaDR2}), an extensive database of astrometric measurements for 
sources up to 90~kiloparsecs (kpc) from the solar system, including
proper motions, radial velocity, and photometry in blue $G_{\rm BP}$, red $G_{\rm RP}$, 
and green $G$ passbands, together with
extinction $A_{\rm G}$ and reddening $E(G_{\rm BP}-G_{\rm RP})$, is now available. 
This unique database provides a refined characterization of
the structure and evolution of the Milky Way
\citep{GaiaGC,Helmi2018, Antoja2018} and could provide
useful data for the search of very young stars in the halo region, beyond
the nominal edge of the disk (15 kpc from the Galactic center, \citealt{Armentrout2017}). In
this work we present our efforts to search for recent star formation in the
outskirts of the Milky Way using \textit{Gaia} DR2 data. The sample selection
is explained in Section \ref{sec:sample_selection}, the kinematics of the
potentially
young candidates are studied in Section \ref{sec:kinematics}, and    a discussion of the results is presented in Section
\ref{sec:discussion}. Finally, the main conclusions of our work are summarized in Section
\ref{sec:conclusions}.

\section{Sample}\label{sec:sample_selection}
\begin{figure*}
  \centering
  \includegraphics[width=\textwidth]{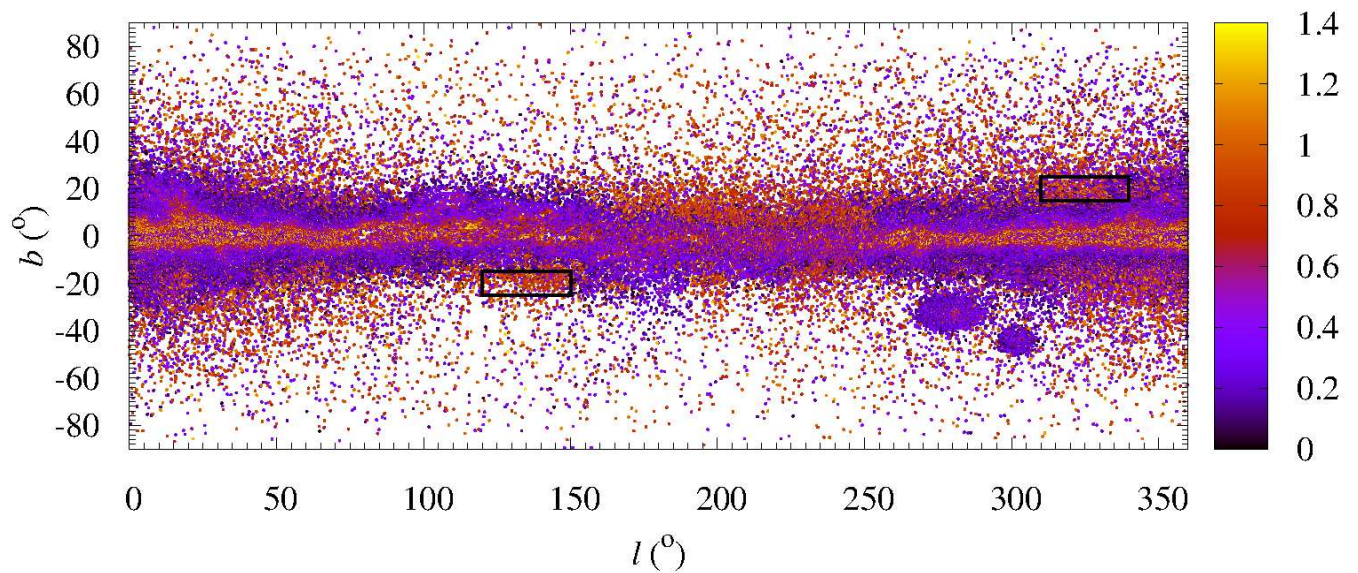}
  \caption{\textbf{Reddening Map for Distant Sources in Galactic Coordinates.} Scatter plot of all \textit{Gaia} DR2 sources with parallaxes between 1/60 and 
                1/30 mas (nominal heliocentric distances between 30 and 60~kpc) in Galactic coordinates. The color-coding corresponds to the 
                value of the reddening, $E(G_{\rm BP}-G_{\rm RP})$. The black rectangles enclose Region 1 (lower left) and Region 4 (upper 
                right).}
  \label{fig:sample_reddening}
\end{figure*}

Young stars are characterized by their strong ultraviolet emission. However,
since in the early phases of stellar evolution they are still embedded
in a dust dense envelope, light at ultraviolet
and optical wavelengths is attenuated. In addition, if they
are far away, the presence of interstellar clouds in the lines of sight
to stars contributes to the overall extinction
and an apparent redder color
is observed. We do not expect to be able to
  resolve stars of the first type in the outskirts
  of the Milky Way because they will be severely extincted
  and their detection unfeasible. Nevertheless, we can
  try to find distant young stars, of a few hundred Myr\footnote{Hereafter, when we  refer to {young stars} we mean stars with an age of $\sim 100-200$ Myr, independently of the evolutionary stage.}, that will be
  naturally reddened by the interstellar material of the disk.

Keeping these ideas in mind, we   searched for
stars with unusually high reddening values in the vast database of \textit{Gaia} DR2.\\

We   selected  all the stars with strictly positive values of
parallax and non-null values
of reddening $E(G_{\rm BP} - G_{\rm RP})$ derived
by the Apsis data processing pipeline \citep{Andrae2018},
resulting in a
catalogue of 87,733,672 stars.

\begin{figure}
  \centering
  \includegraphics[width = \columnwidth]{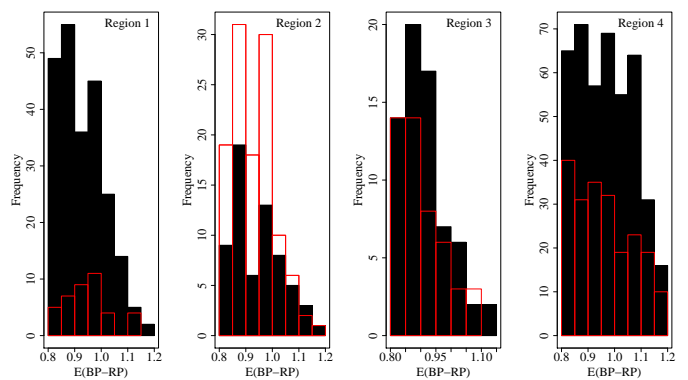}
  \caption{\textbf{Counts of reddened sources.} Distribution of sources
    with reddening values between 0.8 and 1.2 towards the four Regions
    discussed in Section \ref{sec:sample_selection} in black, and
  their mirror   Galactic counterparts (same longitude,
opposite latitude) in red.}
  \label{fig:sym_stats}
\end{figure}

First, we plotted all the stars in slices of a few hundred parsecs
  estimating the distance as the inverse of the parallax and examined
  \textit{Gaia} DR2 data in search of regions with unusual patterns of
  reddening. The most interesting results were found at
  parallax values between 1/60 and 1/30 milliarcseconds (mas),
  which would ideally
  correspond to distances in the range  30-60 kpc, so we focused our attention
  on this subsample.  In addition, these \textit{Gaia} DR2 sources were
  systematically ignored in previous studies because of the extremely
  large parallax errors, usually on the order of the parallax value itself.
  In this way our study can be considered  both a search for a young
  population in the Milky Way halo and an independent assessment of the
  reliability of extreme \textit{Gaia} DR2 measurements. For stars
  with parallax values between 1/60 and 1/30 mas, we
 searched for regions detached from the Galactic
disk (Galactic latitudes $|b| > 15^{o}$) where a moderate number of extincted sources with reddening greater than
0.8 are present. We  
identified four regions that meet these
requirements (see Figure \ref{fig:sample_reddening});
we labeled them  Region 1  $(l, b) = (135^{o}, -20^{o})$,
Region 2 $(l, b) = (195^{o}, -15^{o})$,
Region 3 $(l, b) = (230^{o}, -20^{o})$,
and Region 4 $ (l, b) = (325^{o}, +20^{o})$.
From these four overdensities we  selected
Regions 1 and 4 because they are
significantly redder than their mirror Galactic counterparts (same longitude,
opposite latitude; see Fig. \ref{fig:sym_stats}). \par

Since the parallax errors are extremely large, if we apply
any cut based on them the resulting sample will be practically
nonexistent. On the other hand, we can use the
renormalized unit weight error (RUWE) measurement associated
with each \textit{Gaia} DR2 source to determine
the quality and reliability of the astrometric parameters
\citep[see][for details]{GaiaDR2, RUWE}.
Basically the RUWE value indicates
whether the derived parameters for the source are reliable or if there
are any problems with the astrometric solution. For the stars
in our sample, the RUWE value is less than 1.4 for more than
95\% of the objects, meaning that their astrometric
solution is valid. 
We also  did not apply   the zero-point correction since it is
discouraged by \citet{Lindegren2018} and there are serious
discrepancies in the exact value
(see, {e.g.}, the discussion in \citealt{dFM2019} and references
therein).
In addition, it is not
clear how our regions might be affected, and 
the correction value is  the same order of magnitude
as the parallaxes. However, we were able to confirm independently
that some of them are
actually that far away (see Section \ref{sec:discussion}).\par

We note that
contamination may arise from the Galactic disk but also
from stellar and Galactic streams. Towards Region 1 and
at heliocentric
distances of $\sim$ 18--20 kpc lies the Triangulum-Andromeda (TriAnd)
overdensity \citep{Bergemann2018,Hayes2018}, which according to  recent studies has an origin in the Milky Way
disk, but whose stars have ages ranging from 6 to 10 Gyr \citep{Bergemann2018}; there is no known stellar stream or overdensity associated
with Region 4. In both cases, the peculiar stellar population seems
to be well detached from the bulk of the disk population (see
Fig. \ref{fig:sample_reddening}).

\section{Kinematics}\label{sec:kinematics}
Motivated by the high values of reddening of our stars in
\textit{Gaia} DR2, we 
studied their kinematics in order to search for evidence of
internal coherence. As we want to determine if they were formed
inside the Milky Way or, on the other hand, are part of an
external stellar stream, we   compared the data with
synthetic data from the Galactic model
GALAXIA \citep{Galaxia}, which takes into account the warp and flare of the
disk. \par

\begin{figure}
  \centering
  \includegraphics[width=\columnwidth]{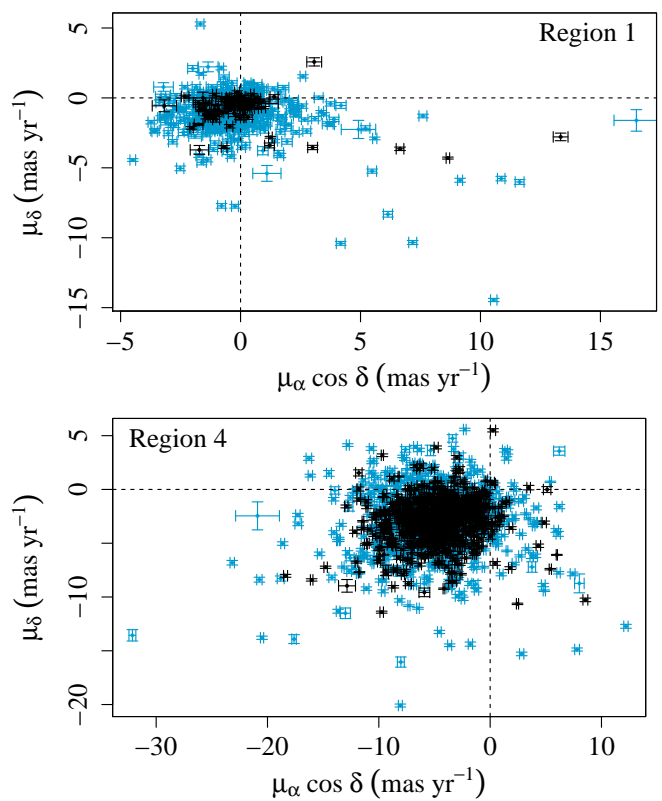}
  \caption{\textbf{Proper Motions.} Proper motions of \textit{Gaia} DR2 sources in Region 1 and
         Region 4. Blue 
                points (376 in Region 1; 741 in Region 4) correspond to sources with values of $G_{\rm BP}-G_{\rm RP} < 0.5$, while black 
                points (83 in Region 1; 690 in Region 4) correspond to sources with 
                $G_{\rm BP}-G_{\rm RP} > 0.5$.}
  \label{fig:proper_motions}
\end{figure}

First, we  studied the proper motions of both regions, which
are displayed in Figure \ref{fig:proper_motions}.
In the analysis, we  distinguished between sources
with color $(G_{\rm BP}- G_{\rm RP}) - E(G_{\rm BP}-G_{\rm RP}) < 0.5 $ (blue points) and $> 0.5$ (black points) since we   expected that any
young stellar candidates would lie within the blue population.
For an A7V star, $B-R = 0.5$ is the turnoff point of a 165 Myr old
  metal-rich cluster, so it makes sense to assign this arbitrary boundary when looking for relatively young populations and it is also consistent with our own definition of youth (see, {e.g.}, \citealt{2005PASP..117.1187H}).
Although
we observe that there is no apparent difference between the
black and blue populations (see Table \ref{tab:pm_statistics}), stars in Region 4 are far more dispersed
and their proper motions are five times higher than those in
Region 1.\par

\begin{table}
  \caption{Median and interquartile range (IQR) values in mas yr$^{-1}$ for the proper motions
  of stars in Regions 1 and 4. }
  \label{tab:pm_statistics}
  \centering
  \begin{tabular}{r|cc|cc}
    \hline \hline
    & \multicolumn{2}{c}{Region 1} & \multicolumn{2}{c}{Region 4} \\
    \hline
    & Blue points & Black points & Blue points  & Black points\\
    \hline
     $\mu_{\alpha}\cos{\delta}$ - median & -0.594  & -0.300 & -5.018 & -4.560 \\
     IQR & 1.772 & 1.277 & 4.977 &  3.496 \\
    $\mu_{\delta}$ - median & -1.029 & -0.551 & -2.607 & -2.715\\
     IQR & 1.569 & 0.927 & 3.593 & 2.421\\
    \hline
  \end{tabular}
\end{table}

\begin{figure}
  \centering
  \includegraphics[width=\columnwidth]{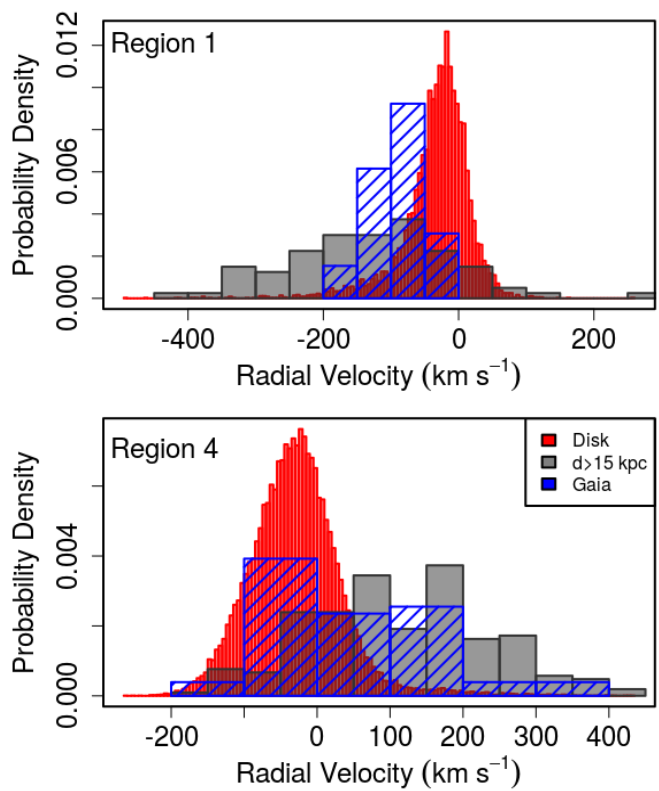}
  \caption{\textbf{Radial Velocities.} Histogram of radial velocities of sources from Regions 1 and  4. Red 
                bars correspond to GALAXIA predictions for the disk of the Milky Way, while gray bars correspond to predictions for the halo; blue dashed bars correspond to probability 
                densities of \textit{Gaia} DR2 data (13 sources in Region 1; 51 in Region 4). The probability density is an amount
                such that the sum of the values
                times the bin size is equal to 1. Bins were computed according to the Freedman-Diaconis rule.}
  \label{fig:radial_velocities}
\end{figure}

The key parameter left to fully determine
the kinematics of the populations is the radial velocity.
Unfortunately, \textit{Gaia} DR2 only provides measurements
for 13 sources in Region 1; for Region 4 we have a statistically
more significant sample including 51 stars. Due to the scarcity of available
data, we  performed a very simple, non-parametric statistical study
based on the median value of radial velocity given by the  GALAXIA model and the
data, and the interquartile range (IQR).\footnote{The IQR is a measure of
  statistical dispersion that indicates the width of the interval where
  50\% of the data are included.}  The results are summarized in
Table \ref{tab:radvel_stats} and complementary
histograms are shown in Figure \ref{fig:radial_velocities}. Although the IQRs are wide, it is clear
that at least Region 1 shows an anomalous distribution in radial
velocities, while Region 4 data could suffer from disk contamination.

\begin{table}
  \caption{Radial velocity statistics (in km s$^{-1}$)
    for the sources in the regions of interest.}
  \label{tab:radvel_stats}
  \centering
  \begin{tabular}{ccccccc}
    \hline \hline
    & \multicolumn{2}{c}{Data} & \multicolumn{2}{c}{Model - disk} & \multicolumn{2}{c}{Model - halo} \\
    & Median & IQR & Median & IQR & Median & IQR\\
    \hline
    Region 1 & -89 & 39  & -26 & 50 & -132 & 149 \\
    Region 4 &  31 & 173 & -31 & 74 &  106 & 164 \\
    \hline
  \end{tabular}
\end{table}

\begin{figure}
      \centering
       \includegraphics[width=\columnwidth]{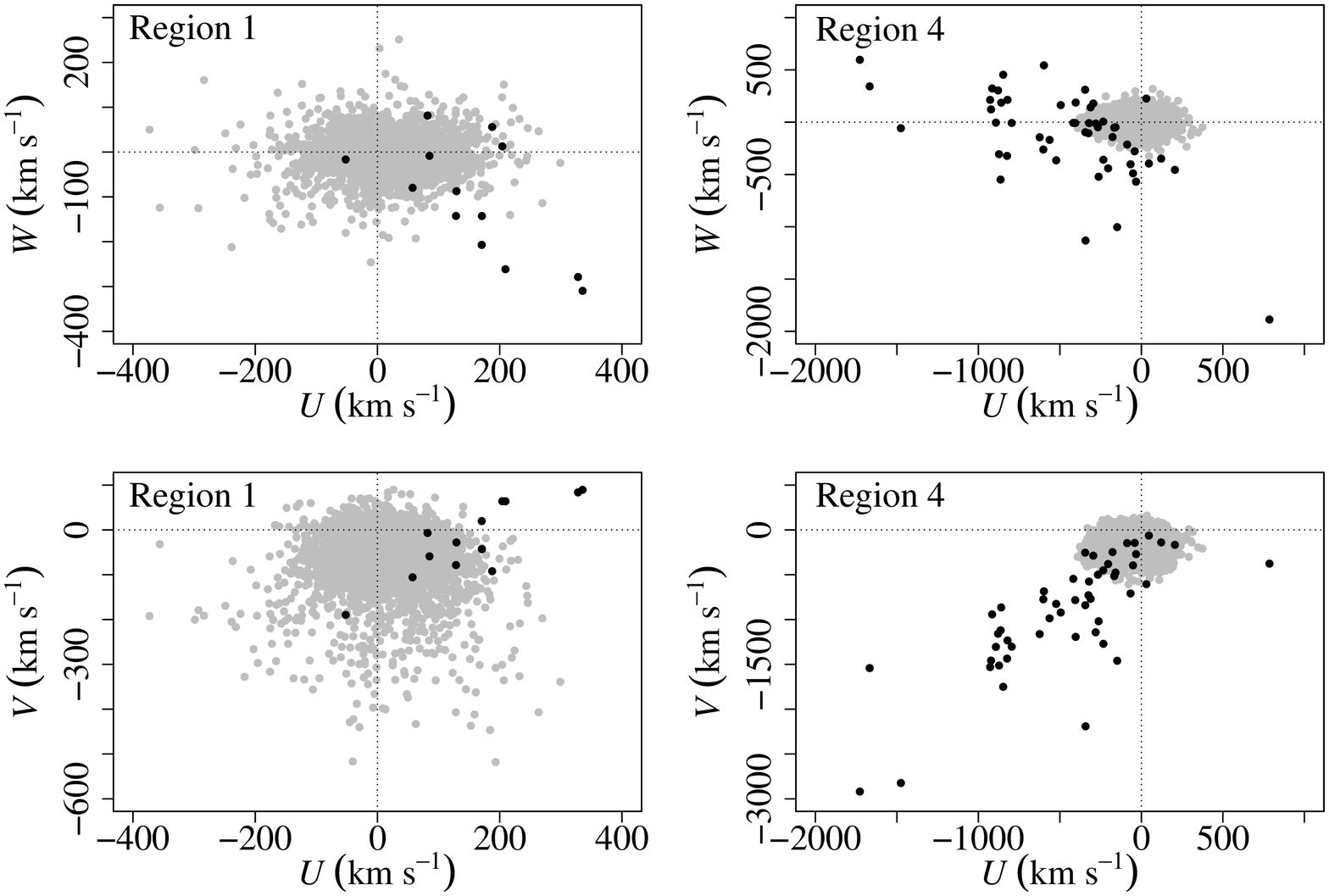}
       \caption{\textbf{Heliocentric velocities.} Heliocentric velocities for sources towards Regions 1 and  4. Gray dots are GALAXIA predictions ($7,065$ in Region 1; $26,031$ in Region 4), black dots are
         \textit{Gaia} DR2 sources with radial velocity data
         (13 in Region 1; 51 in Region 4).} 
       \label{fig:heliocentric_velocities}
\end{figure}

Finally, for the samples with radial velocity, we  computed the
heliocentric Galactic velocities $UVW$ following the standard procedure
 \citep[see, e.g., ][]{Johnson1987}
and compared them with predictions from GALAXIA
(Figure \ref{fig:heliocentric_velocities}); the associated errors were also  computed,
but are not displayed due to the extremely large uncertainties.
From these
plots we interpret that Region 1 data are compatible with a Galactic
origin, perhaps due to a possible contamination arising from TriAnd, while the
trends in Region 4 
indicate that these sources were not born in the place where they
are observed now. What we conclude in view of the very
different kinematics is that these
two regions do not share a common origin.

\begin{figure}
  \centering
  \includegraphics[width=\columnwidth]{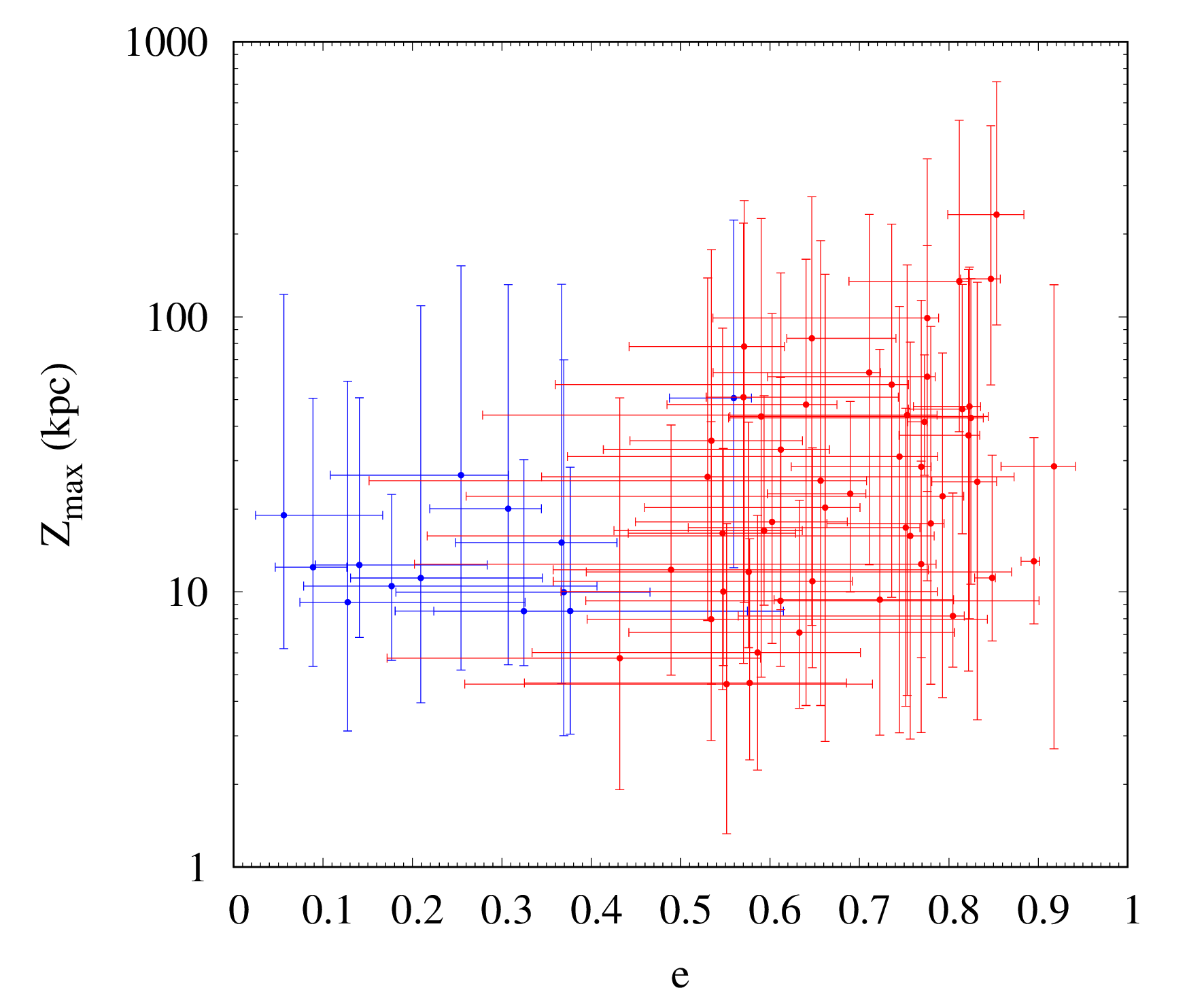}
  \caption{\textbf{Galactic orbits} characteristic of sources in Regions 1 and 4. Maximum height above the plane of the orbit as a function of the eccentricity for relevant
sources in Region 1 (13, in blue) and Region 4 (52, in red). Median values and error bars displaying the 16th and 84th percentiles are shown. }
  \label{galpyresults}
\end{figure}

The Galactic orbits of stars in \textit{Gaia} DR2 with full data sets (i.e., also including radial velocity) can be estimated using Galpy \citep{2015ApJS..216...29B}. 
For those stars in Regions 1 and 4 with complete kinematic information (13 and 52 sources, respectively), we  computed the orbital motions by applying Monte Carlo 
sampling. We  generated 10$^{4}$ instances of the input data set required to calculate one orbit using values and uncertainties from \textit{Gaia}~DR2, then 
extracted relevant parameters such as the maximum value of the $Z$ coordinate ($Z_{\rm max}$) and the eccentricity ($e$). For each source and using the 10$^{4}$ 
orbital realizations, we  computed median values and the 16th and 84th percentiles to estimate the uncertainties, as we describe in the previous sections. The orbits were 
calculated using the gravitational potential recommended in the Galpy documentation, MWPotential2014 \citep{2015ApJS..216...29B}, which includes a bulge, a disk, and a 
dark matter halo component, but neither spiral arms nor giant molecular clouds. Our results are shown in Figure~\ref{galpyresults}. Most sources are fully consistent 
with halo membership, but the samples from Regions 1 and 4 clearly have different provenance. The sample in Region 4, if located as far away as the data suggest, might 
not have an origin in the Milky Way. A possible formation scenario for these
stars would be the collision of a high velocity cloud with the disk \citep{1962ApJ...136..748E}

\section{Discussion}\label{sec:discussion} 
\begin{figure}
  \centering
  \includegraphics[width=\linewidth]{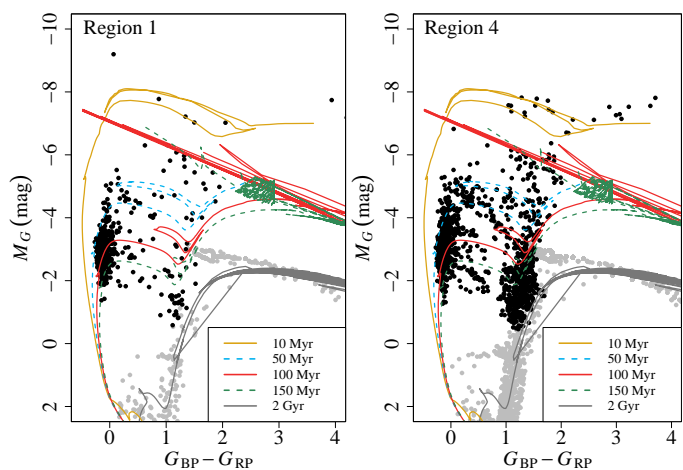}
  \caption{\textbf{Color-magnitude diagrams} for Regions 1  and  4
    for \textit{Gaia} DR2 data,
    corrected from extinction and reddening and with distance
    estimated inverting the nominal parallax; for comparison,
    GALAXIA predictions are shown as gray dots.
    Shown are (from top to bottom)
     PARSEC+COLIBRI PR 16 isochrones with solar metallicity of 10 Myr
    (gold line), 50 Myr (dashed blue line), 100 Myr (solid red line),
    150 Myr (dashed green line), and 2 Gyr (solid gray line). 
  }
  \label{fig:CMD_Gaia}
\end{figure}

\begin{figure}
  \centering
  \includegraphics[width=\linewidth]{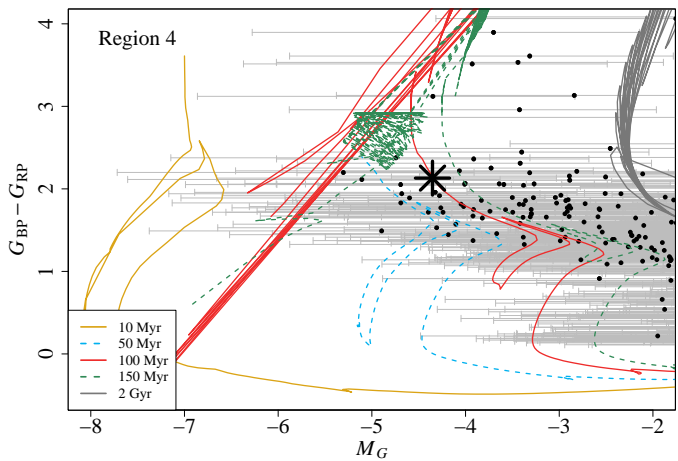}
  \caption{\textbf{Zoom-in of the color-magnitude diagram} for  Region 4
    after applying the Monte Carlo sampling. For better illustration,
    the axes have been rotated and the error bars are displayed in gray. The
    big black star is one of the most promising  young star candidates,   Gaia DR2 6116708946261654272,  because
of the bounded errors
    in absolute magnitude (no additional information in Vizier has been found).
  }
  \label{fig:CMD_MC}
\end{figure}

In the first approach, we  plotted the CM diagram  for Regions 1 and 4, taking
\textit{Gaia} DR2 parallax, extinction, and reddening values; the results are 
displayed in Figure \ref{fig:CMD_Gaia} with some  overlaid PARSEC+COLIBRI PR16
isochrones of solar metallicity \citep{Marigo2017,2012MNRAS.427..127B} and GALAXIA
predictions. In both cases the
data suggest the presence of a blue population that is evolving
towards the
red branch of the diagram, much younger than the current
$\sim 2$ Gyr population already known and predicted by the models. However, while the errors in apparent magnitudes
are not significant, the errors
in the parallax are considerable. The most suggested line of
action in these cases is to apply Bayesian inference to predict distances
from the parallaxes \citep{Luri2018,2018AJ....156...58B}, but
due to the large uncertainties, the posterior distribution will be
strongly affected by the prior, and hence the predicted distances
will be shorter than they really are. Instead, we 
followed the approach outlined by \citet{dFM2019}
and  estimated the heliocentric and Galactocentric
distances via Monte Carlo
sampling, assuming a distance to the Galactic Center of
$d = 8.18$ kpc \citep{Gravity2019}; for each source, we 
performed $10^{5}$ Monte Carlo simulations. Since
\textit{Gaia} DR2 values for extinction and reddening were
derived neglecting the parallax uncertainties \citep{Andrae2018}, we 
determined the absolute magnitudes solely from our Monte Carlo parallaxes.
Figure \ref{fig:CMD_MC} shows the sources of Region 4 for which the
Monte Carlo estimates are compatible with a young age, although it is clear that
  the errors in the parallaxes are so large that we cannot be
  certain about the values. The most promising candidate in Region 4 is
  Gaia DR2 6116708946261654272, with lower bar error in
  absolute magnitude $M_{G}^{-} < -3$ and $16{th}$
  Galactocentric distance percentile\footnote{We  computed
  the 16th ($p_{16}$) and 84th ($p_{84}$) percentiles following Apsis convention. They can be
  considered  asymmetric error bars for the median estimates.} greater than 15 kpc. Thus, we  performed an extensive search in
  both Vizier\footnote{All the matches in Vizier were
    carried out in a search radius of 3''.} \citep{2000A&AS..143...23O}
  and SIMBAD \citep{2000A&AS..143....9W}
  looking for any available information about the samples that could shed some
  light on the question of their age. What we  found is that in general,
  there is not much information about sources in Region 4, which we
  ascribe to
  their location in the southern celestial hemisphere, while for sources
  in Region 1 there are some LAMOST DR4\footnote{\href{http://dr4.lamost.org/}{http://dr4.lamost.org/}}  \citep{LAMOSTDR4} spectra available that we  used
  for the purpose of
  estimating spectroscopic parallaxes. Eventually, we   found one
  star, Gaia DR2 375075920547984000, which is classified as an A2IV by
  LAMOST DR4. We
   estimated the Galactocentric distance to the source
  spectroscopically and via Monte Carlo sampling, and the results are
  compatible. For the spectroscopic estimate, we took YZ
  Cassiopeiae as an A2IV reference star, with a visual magnitude
  $M_{V} = 0.251$ mag
  \citep{2008AN....329..835B},
  and we obtained a heliocentric distance of 21.2 kpc with an estimated
  error of 0.9 kpc\footnote{We  used LAMOST $g$ and $r$ magnitudes for the
    estimation of the V magnitude via the relationship
    $V = r + 0.44(g-r)-0.02$, considering an error of 0.05 mag
  for each magnitude.}, which
  corresponds to a Galactocentric distance of 26.3 kpc for this object;
  the Monte Carlo estimates are compatible with a
  median Galactocentric distance of 17.5 kpc, and
  percentiles $p_{16}$ = 11.1 kpc and $p_{84}$ = 49.6 kpc.
  This star is also included in the Starhorse catalog recently
  published by \cite{starhorse} although the distance estimate is not
  very reliable; the resulting distance probability density function
  is very broad. Nevertheless, their Galactocentric distance is
  also compatible with our results, with a value of 26.3 kpc. In summary,
  we have an A2IV star located farther than 15 kpc from the Galactic center.
  This star is very likely to be young since an A2 star spends little time
  in the subgiant stage, and in consequence there should be more
  young stars with it, unless it is a runaway star (see, e.g., \citealt{2000ApJ...544L.133H}); no conclusions   can be drawn without a better characterization of the star
  (radial velocity, reddening, and precise parallax). \par
To determine whether there are more  young star candidates like the one 
  just discussed, we need to be sure that \textit{Gaia} DR2
  nominal parallaxes
  are valid despite the large errors. Our search in Vizier has reported
  two more encouraging results. Gaia DR2 396558526625729152 is classified
  in LAMOST DR4 as an A2V that gives a spectroscopic Galactocentric
  distance of 17.4 kpc (taking $M_{V} = 1.3$ mag, \citealt{2000asqu.book..381D}), and
  Gaia DR2 338768431691725824 is an RR Lyrae star \citep{Sesar2017} with
  a distance modulus of 16.31 mag. Although the RR Lyrae is not a young
  star, both of them are well beyond the nominal edge of the
  Galactic disk. We   are thus convinced that \textit{Gaia} DR2 parallaxes
  for distant sources are more reliable than previously thought.

However, there could be some contamination arising from TriAnd
  suggested by the presence of two stars in Region 1, Gaia DR2
  374544482769997056 and Gaia DR2 330307728370143360, that
  were previously   identified by \citet{Sheffield2014} as
  bona fide members of TriAnd. In addition,
  an old population is clearly revealed by LAMOST: Gaia DR2 339458031639126016
  is classified as a M1 star, and Gaia DR2 388342971445046272 as a K7.
 The list of the discussed
stars is shown in Table \ref{tab:stars}.\par

\begin{table*}
\caption{Stars with additional data discussed in Section \ref{sec:discussion}.}             
\label{tab:stars}
\centering                          
\begin{tabular}{c c c c}        
\hline\hline                 
Star & Other IDs & Notes & Reference \\
\hline                      
Gaia DR2 375075920547984000 & LAMOST DR4 197116019  & A2IV & \citet{LAMOSTDR4} \\
Gaia DR2 396558526625729152 & LAMOST DR4 370404083 & A2V & \citet{LAMOSTDR4} \\
Gaia DR2 338768431691725824 &                  &  RR Lyrae & \citet{Sesar2017}\\
Gaia DR2 374544482769997056 &                  & TriAnd member & \citet{Sheffield2014} \\
Gaia DR2 330307728370143360 &            & TriAnd member & \citet{Sheffield2014} \\
Gaia DR2 339458031639126016 & LAMOST DR4 191606221  & M1 & \citet{LAMOSTDR4} \\
Gaia DR2 388342971445046272 & LAMOST DR4 164415092  & K7 & \citet{LAMOSTDR4} \\

\hline                                  
\end{tabular}
\end{table*}

\section{Conclusions}\label{sec:conclusions}
If we take into account the analysis presented in the previous
section, we are convinced that nominal \textit{Gaia} DR2 parallaxes
may be valid even for very distant sources ($d > 15$ kpc), since our
Monte Carlo estimated values are always compatible with the
few spectroscopic
estimates available. Nevertheless, the topic of the age of the
sources is still debatable. In view of the corrected data
shown in the CMD in Figure \ref{fig:CMD_MC},
and assuming (based on the few confirmed cases) that most of the
stars are really that far away, we conclude that within the samples there
must be some young stars,  the most
promising candidate being Gaia DR2 375075920547984000. On
the other hand, although the stars are likely
  to be distant, they are not located farther than 30 kpc,
  meaning that the errors in the parallax are not negligible.
  Hence, reddening and extinction values
derived by Apsis are likely to change, and we have to be
 cautious with respect to the CMD and
the results that can be drawn from them.
We are hopeful that  with the arrival of \textit{Gaia} DR3 we will be
able to select a reliable sample of candidates in both regions
to be studied in more detail.
  
\begin{acknowledgements}
  We want to thank an anonymous referee for a careful revision that
  has greatly improved this manuscript.
  This work has been partly funded through grants ESP2015-68908-R and ESP2017-87813-R. LBA acknowledges Complutense University of Madrid and Banco Santander for the grant ``Personal Investigaci\'on en Formaci\'on CT17/17-CT18/17".
    The authors acknowledge Javier Y\'a\~nez for useful discussions on statistical strategies.
         In preparation of this article, we made use of the NASA Astrophysics Data System and the ASTRO-PH e-print server. This research has 
         made use of the  SIMBAD and VizieR  databases operated at CDS, Strasbourg, France. This work has made use of data from the European 
         Space Agency (ESA) mission {\it Gaia} (\url{https://www.cosmos.esa.int/gaia}), processed by the {\it Gaia} Data Processing and 
         Analysis Consortium (DPAC, \url{https://www.cosmos.esa.int/web/gaia/dpac/consortium}). Funding for the DPAC has been provided by 
         national institutions, in particular the institutions participating in the {\it Gaia} Multilateral Agreement. This research made use of Astropy, a community-developed core Python package for Astronomy \citep{astropy2013, astropy2018}. Some of the plots included in this article as well as the statistical
         treatment of the data have been performed with R \citep{Rmanual}.
\end{acknowledgements}

%
%
\bibliographystyle{aa}
\bibliography{references}
\end{document}